\overfullrule=0pt
\input psfig.sty

\bigskip
\font\lgh=cmbx10 scaled \magstep2

\magnification=1100
\voffset 0.5truein

\def\mediumspace{\baselineskip 18pt \lineskip 6pt \parskip 3pt plus 5pt}

%
\mediumspace
\raggedbottom
%
%
\def\deg{\ifmmode^\circ\else$^\circ$\fi}

\def\simlt{\hbox{ \rlap{\raise 0.425ex\hbox{$<$}}\lower
0.65ex\hbox{$\sim$} }}
\def\ltorder{\hbox{ \rlap{\raise 0.425ex\hbox{$<$}}\lower
0.65ex\hbox{$\sim$}}}
\def\simgt{\hbox{ \rlap{\raise 0.425ex\hbox{$>$}}\lower
0.65ex\hbox{$\sim$} }}
\def\gtorder{\hbox{ \rlap{\raise 0.425ex\hbox{$>$}}\lower
0.65ex\hbox{$\sim$}}}
\def\sles{\lower2pt\hbox{$\buildrel {\scriptstyle <}
   \over {\scriptstyle\sim}$}}
\def\sgreat{\lower2pt\hbox{$\buildrel {\scriptstyle >}
   \over {\scriptstyle\sim}$}}

\def\qquad{\quad\quad}

\def\gsim{\lower 2pt \hbox{$\, \buildrel {\scriptstyle >}\over
{\scriptstyle \sim}\,$}}
\def\lsim{\lower 2pt \hbox{$\, \buildrel {\scriptstyle <}\over
{\scriptstyle \sim}\,$}}

%
%
\def\refto#1{$^{#1}$}           
\def\ref#1{Ref.~#1}                     
\gdef\refis#1{\item{#1.\ }}                     
\def\beginparmode{\endmode
  \begingroup \def\endmode{\par\endgroup}}
\let\endmode=\par
\def\body{\beginparmode}
\def\head#1{                    
  \goodbreak\vskip 0.5truein    
  {\centerline{\bf{#1}}\par}
   \nobreak\vskip 0.25truein\nobreak}
\def\references                 
  {\head{References}            
   \beginparmode
   \frenchspacing \parindent=0pt \leftskip=1truecm
   \parskip=8pt plus 3pt \everypar{\hangindent=\parindent}}
\def\endreferences{\body}

\input reforder.tex    

%
%
\centerline{\lgh On the Nature of the X-ray Emission from}
\centerline{\lgh the Galactic Center Region}
\bigskip
\centerline{\bf Q. D. Wang$^*$, E. V. Gotthelf$^\dagger$, \&
 C. C. Lang$^*$}
$^*$ Department of Astronomy, University of Massachusetts, Amherst, MA 01003,
USA\par 
$^\dagger$ Columbia Astrophysics Laboratory, Columbia University, New York, NY
10027, USA
\medskip

\noindent

{\bf The origin of the X-ray emission from the central region of the Galaxy
has remained a mystery\refto{Koy96}$^,$\refto{Sid99}$^,$\refto{Val00}$^,$\refto{Tan00}, despite extensive study over the past two decades. A fundamental 
question is the relative contribution of the point-source
and diffuse components of this emission, which is critical to understanding
the high-energy phenomena and processes unique
to this Galactic nuclear environment.
Here, we report on results from a large-scale imaging survey of the Galactic 
center with sufficient spatial resolution to allow a clean separation of
the two components. The He-like Fe K$\alpha$ emission, previously
attributed to the diffuse emission\refto{Koy96}, is found largely due to the
discrete X-ray source population. The number and spectrum of such sources
indicate the presence of numerous accreting white dwarfs, neutron stars, 
and/or black holes in the region. The diffuse X-ray emission 
dominates over the contribution from the faint
discrete sources and is shown to be associated with distinct
interstellar structures observed at radio and mid-infrared wavelengths,
suggesting that it arises from the recent formation of massive stars.}

We have carried out a systematic X-ray survey of the Galactic Central (GC)
region at arcsecond scales with the {\it Chandra} X-ray Observatory\refto{Wang01}. Figure 1 presents the image of the X-ray emission arising 
from the GC region.  
The figure, though optimized to reveal low surface brightness features, shows
many discrete sources. Our preliminary source detection analysis results in $\sim$
1000 discrete sources in the field with a signal-to-noise of $\simgt3\sigma$. Less than 20 of these sources are previously known objects, most of which are 
bright X-ray binaries\refto{Sid01}$^,$\refto{Sak01}; examples include the two
bright accreting X-ray binaries, 1E 1743.1-2843 and 1E 1740.7-2942. Based on 
a comparison with the source
density in a relatively blank region of the Galactic plane\refto{Eb01},
we estimate that up to half of our newly detected sources could be luminous
background active galactic nuclei, the X-ray radiation from which is highly 
attenuated by the Galactic interstellar absorption. 
The X-ray absorption also varies across the field, affecting the surface
brightness distribution, particularly in the $1-3$ keV energy band.  The majority
of the detected sources are in the energy range of $2-10$ keV with
a luminosity range of $10^{32} - 10^{35} {\rm~ergs~s^{-1}}$ at the
distance of the GC.

We consider the nature of the large number of newly-detected X-ray
sources by examining their composite spectral signature and compare this 
to the diffuse emission (Figure 2). 
The accumulated source spectrum (excluding the two brightest ones)
within the central region shows a
distinct emission feature centered at $\sim 6.7$~keV with a Gaussian
width of $\sim 0.09$~keV, which agrees with previous
measurements\refto{Tan00}. This spectral feature represents
the He-like Fe K$\alpha$-line. Modest contributions from the 
$\sim 7.0$-keV H-like Fe K$\alpha$-line and the 6.4-keV fluorescent line
of neutral to moderately ionized Fe atoms are also present in the spectrum. 
These emission lines are characteristic of X-ray binaries 
containing white dwarfs, neutron
stars, or  black holes, especially at relatively quiescent 
states\refto{Bar00}$^,$\refto{Feng01}. This X-ray binary
explanation is also consistent with the otherwise flat and relatively
featureless shape of the spectrum. But other possibilities cannot be ruled out.
The narrow emission lines (at $\sim$ 2.4 keV and possibly at other energies) 
are likely due to a relatively small contribution from massive stars. 
The X-ray spectrum of the luminous Arches star cluster\refto{Yusef01}, for example, shows distinct narrow lines and is
considerably softer than the mean source spectrum shown in Figure 2. The upturn
of the source spectrum below $\sim 2$ keV is a result of the contribution from
relatively nearby sources, which suffers less amounts of soft X-ray absorption by the interstellar
medium than the X-ray radiation from the GC.

Our survey also shows that large amounts of 
diffuse X-ray emitting material are distributed  asymmetrically
around the GC and are particularly concentrated in the region about 
$10^\prime$ to the lower left of Sgr A$^*$ (\ref{B01}; Figure 2). 
Several additional large-scale diffuse X-ray features are present in the 
soft band ($1-3$ keV) around the Sgr A$^*$, some of which apparently 
extend into high Galactic latitude regions outside the field of view of the
X-ray image. There are also prominent features associated with the Sgr C complex 
and near the Galactic microquasar 1E 1740.7-2942. 
The count rate ratio of the sources to the enhanced diffuse component
in the two spectra (Figure 2) is $\sim 0.25$ in the $2-10$ keV band. 
Even in the regions of the deepest exposure ($\sim 120$ ks in the Sgr B2 region,
including an archival observation\refto{Mur01}), where the source detection
limit reaches $\sim 8 \times 10^{31} {\rm~ergs~s^{-1}}$, the diffuse
emission dominates. There is no known X-ray source population that might
contribute substantially below this source detection limit.

The He-like Fe K$\alpha$-line (at $\sim 6.7$ keV) is not as prominent in 
the diffuse emission spectrum as in the discrete
source spectrum. The ubiquitous and strong presence of this line in
previous X-ray spectra of the GC region\refto{Koy96} is partly due to
discrete sources. The reduced strength of the He-like 
Fe line no longer requires the presence of large amounts of $\sim 10^8$ K gas,
which are very difficult to explain physically\refto{Tan00}.
The weaker He-like Fe line as well as the prominent ion lines at 
lower energies (e.g., Si XIII K$\alpha$ and S XV K$\alpha$;  Figure 2) 
are consistent with an optically-thin thermal plasma with a characteristic
temperature of $\sim 10^7$ K, comparable to that typically found in 
young supernova remnants. Such plasma still significantly 
contributes to the 6.7-keV line. One example is Sgr A East, 
most likely a young supernova remnant, which has an X-ray spectrum characterized by 
a thermal plasma with a temperature of $\sim 2$ keV and shows a strong 
6.7-keV emission line\refto{Mae01}. 

Our spectral results also suggest that nonthermal processes contribute
significantly to the diffuse X-ray emission, especially at the higher energies. A recent study
has demonstrated that the inclusion of a non-thermal component in modeling the X-ray
background spectrum from the Galactic plane and toward the GC significantly reduces the
required characteristic temperature of the plasma component\refto{Val00}. But the nature
and origin of this 
contribution remain unclear. Nearly half of our
detected diffuse emission in the 5-8 keV band is due to the Fe 6.4-keV line
(Figure 2), which results from the filling of neutral to moderately ionized 
Fe K shell vacancy due either to ionization by hard
X-ray radiation ($>$ 7.1 keV) or to collision with non-relativistic cosmic
rays\refto{Val00}. 
We find that the distribution of the line emission
is indeed generally correlated with lumpy dense
molecular material\refto{Jac96}$^,$\refto{Price01}. However, currently there is
not a sufficient population of bright X-ray sources in the GC region to produce
the 6.4-keV line fluorescence\refto{Mur01}$^,$\refto{From01}. 
One possibility might be that the luminosity of X-ray sources (e.g., the GC
supermassive black hole) varies greatly and that the currently low luminosity is
abnormal. If the luminosity averaged over the light crossing time of the region
(a few hundred years) is several orders of magnitude higher, much of the 5-8 keV
band diffuse emission may then be explained as the past 
point-like emission scattered/fluoresced off 
molecular clouds found in the region\refto{Tsuboi97}$^,$\refto{Jac96}.

We also find that the continuum emission in the 4-6 keV range is substantially
more uniformly distributed along the central Galactic plane than both
the 6.4-keV line and the emission in the lower energy bands. This continuum
emission may be induced partly by nonthermal electrons.
However, we find little correlation between diffuse X-ray and radio features in
the GC region. Of the eight prominent nonthermal filaments (NTFs) known, only
one has a direct X-ray counterpart: G359.54+0.18 (\ref{Yusef97}; marked as
``X-ray Thread'' in Figure 1). In particular, the X-ray emission is not 
correlated with the most prominent NTFs in the Radio Arc or with the
mid-infrared shells/arcs (e.g., Figure 3). This suggests that the bulk of the
emission cannot be due to the inverse Compton scattering of the cosmic 
microwave background or the interstellar infrared radiation off relativistic
electrons\refto{Boggs00}. But bremsstrahlung and Fe K shell vacancy from
non-relativistic cosmic-ray electrons ($\simlt$ 1 MeV) remain a
possibility\refto{Val00}. 

Such multiwavelength comparisons have provided us with a 
new perspective of the interplay between various stellar and interstellar
components in the GC region. The enhanced diffuse X-ray emission is globally
correlated with massive star forming regions, which include the GC, Arches, and
Quintuplet clusters\refto{Figer99}. In particular, the Quintuplet cluster
is thought to be responsible for producing distinct shell-like structures as
seen in radio and mid-infrared (Figure 3). These structures, together with the enclosed
diffuse X-ray-emitting materials, are likely a product of
mechanical energy release from massive stars in form of supernova 
explosions and fast stellar winds.

It is interesting to compare the massive star forming regions 
of the GC with the 30 Doradus nebula. The collective energy releases in 
the two regions are comparable\refto{Figer99}$^,$\refto{Wang99}. 
With an overall extent of about 300 pc, the 30 Dor nebula consists of blisters 
of X-ray-emitting gas enclosed in photon-ionized loops and 
shells\refto{Wang99}. But the bulk
of the X-ray emission arises in gas with a characteristic temperature 
$\simlt 10^7$ K, and the gas apparently cools off with increasing distance 
from the central cluster R136. However, such gas in the GC region is 
not traced 
by X-rays observed in our {\it Chandra} survey, because of the heavy 
foreground absorption ($\sim 10^{23} {\rm~cm^{-2}}$). The strongly 
enhanced hard X-ray emission ($\simgt 4$ keV) is thus very unique to the GC 
environment. The expansion and outflow of the high-pressure and buoyant 
plasma/cosmic rays may be responsible for the large-scale, vertical diffuse 
radio and soft X-ray features observed above and below this region of the 
Galaxy\refto{Almy01}$^,$\refto{Sofue00}.
\bigskip


\refis{Almy01} Almy, R. C., McCammon, D. Digel, S.W., Bronfman, L., \& May, J. Distance Limits on the Bright X-Ray Emission Toward the Galactic Center: Evidence for a Very Hot Interstellar Medium in the Galactic X-Ray Bulge. {\it Astrophys. J.}, {\bf 545}, 290-300 (2000)

\refis{B01} Baganoff, F., et al. Rapid X-ray Flaring from the Direction of the Supermassive Black Hole at the Galactic Centre. {\it Nature}, {\bf 413}, 45-48 (2001)

\refis{Bar00} Barret, D., et al. Hard X-ray Emission from Low-Mass X-ray Binaries. {\it Astrophys. J.}, {\bf 533}, 329-351 (2000)

\refis{Boggs00} Boggs, S.E., et al. Diffuse Galactic Soft Gamma Ray Emission. {\it Astrophys. J.}, {\bf 544}, 320-329 (2000)

\refis{Eb01} Ebisawa, K., et al. Origin of the Hard X-ray Emission from the Galactic Plane. {\it Science}, {\bf 293}, 1633-1635 (2001)

\refis{Feng01} Feng, Y. X., et al. Evolution of Iron K$\alpha$: Line Emission in the Black Hole Candidate GX339-4. {\it Astrophys. J.}, {\bf 553}, 394-398 (2001)

\refis{Figer99} Figer, D., et al. Hubble Space Telescope/NICMOS Observations of Massive Stellar Clusters near the Galactic Center. {\it Astrophys. J.}, {\bf 525}, 750-758 (1999)

\refis{From01} Fromerth, M.J., Melia, F., \& Leahy, D.A. A Monte Carlo Study of the 6.4 keV Emission at the Galactic Center. {\it Astrophys. J.}, {\bf 547}, 129-132 (2001)

\refis{Jac96} Jackson, J.M., Heyer, M. H., Paglione, T., \& Bolatto, A. HCN and CO in the Central 630 Parsecs of the Galaxy. {\it Astrophys. J.}, {\bf 456}, L91-L95 (1996)

\refis{Koy96} Koyama, K., et al. ASCA View of Our Galactic Center: Remains of  Past Activities in X-Rays? {\it Publ. Astron. Soc. Japan}, {\bf 48}, 249-255 (1996)


\refis{Mae01} Maeda, Y., et al. A Chandra Study of Sgr A East: A Supernova Remnant Regulating The Activity of Our Galactic Center? {\it Astrophys. J.}, in press, (2001)

\refis{Mur01} Murakami, H., Koyama, K., \& Maeda,  Y., Chandra Observations of
Diffuse X-Rays from the Sagittarius B2 Cloud. {\it Astrophys. J.}, {\bf 558},
687-692 (2001)

\refis{Price01} Price, S. D., et al. Midcourse Space Experiment of the Galactic
Plane. {\it Astrophys. J.}, {\bf 121}, 2819-2842 (2001)

\refis{Sak01} Sakano, M., et al. ASCA X-ray Source Catalogue in the Galactic
Center Region. {\it Astrophys. J. Sup.}, in press, (astro-ph/0108376) (2001)

\refis{Sid01} Sidoli, L., Belloni, T., \& Mereghetti, S. A Catalogue of Soft
X-ray Sources in the Galactic Center Region., {\it Astron. Astrophys.}, {\bf 368}, 835-844 (2001)

\refis{Sid99} Sidoli, L., \& Mereghetti, S. The X-ray Diffuse Emission from the
Galactic Center. {\it Astro. Astrophys.}, {\bf 349}, L49-L52 (1999)

\refis{Sofue00} Sofue, Y. Bipolar Hypershell Galactic Center Starburst Model:
Further Evidence from ROSAT Data and New Radio and X-Ray Simulations. {\it
Astrophys. J.}, {\bf 540}, 224-235 (2000)

\refis{Tan00} Tanaka, Y., Koyama, K., Maeda, Y., \& Sonobe, T. Unusual Properties of X-Ray Emission near the Galactic Center. {\it Publ. Astron. Soc. Japan}, {\bf 52}, L25-30 (2000)

\refis{Towns00} Townsley, L. K., Broos, P.S., Garmire, G., \& Nousek, J. Mitigating Charge Transfer Inefficiency in the Chandra X-Ray Observatory Advanced CCD Imaging Spectrometer. {\it Astrophys. J}., {\bf 534}, L139-L142 (2000)

\refis{Tsuboi97} Tsuboi, M., Ukita, N., \& Handa, T. An Expanding Shell-Like Molecular Cloud near the Galactic Center Arc. {\it Astrophys. J.}, {\bf 481}, 263-266 (1997)

\refis{Val00} Valinia, A., et al. On the Origin of the Iron K Line in the Spectrum of the Galactic X-Ray Background. {\it Astrophys. J.}, {\bf 543}, 733-739 (2000)

\refis{Wang99} Wang, Q. D. Structure and Evolution of Hot Gas in 30 DOR. {\it Astrophys. J.}, {\bf 510}, L139-L143 (1999)

\refis{Wang01} Wang, Q. D., Gotthelf, E. V., Lang, C., \& Jagodzinski, F. UMass/Columbia Chandra X-ray Survey of the Galactic Center Region. {\it Astrophys. J.}, submitted (2001)


\refis{Yusef84} Yusef-Zadeh, F., Morris, M., \& Chance, D. Nature, {\bf 310}, 557-561 (1984)

\refis{Yusef97} Yusef-Zadeh, F., Wardle, M., \& Parastaran, P. The Nature of the Faraday Screen toward the Galactic Center Nonthermal Filament G359.54+0.18. {\it Astrophys. J.}, {\bf 475}, 119-122 (1997)

\refis{Yusef01} Yusef-Zadeh, F., et al., Detection of X-ray Emission from the Arches Cluster near the Galactic Center. {\it Astrophys. J.}, in press, astro-ph/0108174 (2001)

\bigskip
\centerline{\bf REFERENCES AND NOTES}
\medskip

\endreferences

\vfill\eject

\noindent{\bf ACKNOWLEDGMENT.}  We thank F. Jagodzinski for his assistance in 
the data calibration and L. Townsley for helping with the CTI corrections. 

%
%

{\bf Figure 1} --- Mosaic image of the Galactic center region. 
This image covers a $\sim 2 \times 0.8$ square degree band 
in the Galactic coordinates and is centered at $l^{II}, b^{II} = -0.1^\circ,
0^\circ$, roughly the location of the Sgr A Complex. The three 
energy bands are $1-3$ keV (shown in red), $3-5$ keV (green), and $5-8$ keV
(blue). The data consist of 30 separate pointings and were acquired during July
2001 with the front-illuminated Advanced CCD Imaging Spectrometer (ACIS-I).  The
spatial resolution ranges from $\sim 0.5"$ on-axis to $\sim 5"$ at the near edge
of the CCD and to $\sim 10"$ at the diagonal edge. This image
is apaptively smoothed with a signal-to-noise of 3. The intensity is 
plotted logrithmically to emphasize low
surface brightness features. Standard imaging calibration and
exposure-correction have been applied as well as corrections for the charge
transfer inefficiency effects\refto{Wang01}$^,$\refto{Towns00}.

{\bf Figure 2} --- Comparison of accumulated point 
source (lower) and diffuse emission (upper) spectra. 
These two spectra are extracted from an ellipse (with the major
and minor axis equal to 50$^\prime$ and $12^\prime$), 
centered at the Sgr A$^*$ and oriented along the Galactic
plane. Regions around two brightest sources 
(1E 1740.7-2942 and 1E 1743.1-2843) are excluded to minimize the spectral 
pile-up problem. The total count rate from these sources is comparable 
to that of the diffuse emission. For each spectrum, we derive an average 
auxiliary response, weighted by an image of selected events in the detector
coordinates. The response is used for the channel energy scaling and
for characterizing the spectral lines.
A diffuse background, obtained in the outer field and 
normalized in both exposure and area, is subtracted from the two spectra.

{\bf Figure 3} ---  A multiwavelength close-up of the recent massive star 
forming region near the GC. The color image, plotted also in the standard
Galactic coordinates, 
is a composite of 20 cm radio continuum 
(red; \ref{Yusef84}); $25\mu$m mid-infrared (green; \ref{Price01}); 
and 6.4 keV line emission (blue).
\medskip
\noindent\centerline{ {\hfil\hfil
\psfig{figure=figure2.ps,width=14cm,angle=270,clip=}
\hfil\hfil}}

\vfill\eject

\endmode
\end

\noindent\centerline{ {\hfil\hfil
\psfig{figure=f1.ps,width=9cm,angle=0,clip=}
\hfil\hfil}}

\medskip
\noindent\centerline{ {\hfil\hfil
\psfig{figure=f2.ps,width=14cm,angle=270,clip=}
\hfil\hfil}}

\medskip
\noindent\centerline{ {\hfil\hfil
\psfig{figure=f4.ps,width=14cm,angle=90,clip=}
\hfil\hfil}}